\documentclass[a4paper]{IEEEtran}
\usepackage[T1]{fontenc}
\usepackage[latin9]{inputenc}
\usepackage{color}
\usepackage{float}
\usepackage{amsmath}
\usepackage{amsthm}
\usepackage{amssymb}
\usepackage{stackrel}
\usepackage{graphicx}
\usepackage{esint}
\usepackage[unicode=true,
 bookmarks=false,
 breaklinks=false,pdfborder={0 0 1},backref=false,colorlinks=false]
 {hyperref}

\makeatletter

\pdfpageheight\paperheight
\pdfpagewidth\paperwidth

\floatstyle{ruled}
\newfloat{algorithm}{tbp}{loa}
\providecommand{\algorithmname}{Algorithm}
\floatname{algorithm}{\protect\algorithmname}

\theoremstyle{plain}
\newtheorem{thm}{\protect\theoremname}
\theoremstyle{remark}
\newtheorem{rem}[thm]{\protect\remarkname}
\newenvironment{lyxlist}[1]
{\begin{list}{}
{\settowidth{\labelwidth}{#1}
 \setlength{\leftmargin}{\labelwidth}
 \addtolength{\leftmargin}{\labelsep}
 }}
{\end{list}}
\theoremstyle{definition}
\newtheorem{defn}[thm]{\protect\definitionname}

\usepackage{xcolor}
\usepackage{pdfcolmk}
\usepackage{units}

\usepackage{epsfig}
\usepackage{float}
\usepackage{subfigure}
\usepackage{subfloat}
\usepackage{cite}

\usepackage{miniplot}

\allowdisplaybreaks

\PassOptionsToPackage{normalem}{ulem}
\usepackage{ulem}

  \pdfpageheight\paperheight
\pdfpagewidth\paperwidth

\renewcommand{\thesubsubsection}{\arabic{subsubsection}}

\makeatother

\providecommand{\definitionname}{Definition}
\providecommand{\remarkname}{Remark}
\providecommand{\theoremname}{Theorem}

\begin{document}

\title{A Topological Approach to Secure Message Dissemination in Vehicular
Networks}

\author{Jieqiong Chen, \IEEEmembership{Student Member, IEEE}, Guoqiang Mao,
\IEEEmembership{Fellow, IEEE}, Changle Li, \IEEEmembership{Senior Member, IEEE},\\
 Degan Zhang, \IEEEmembership{Senior Member, IEEE}}
\maketitle
\begin{abstract}
Secure message dissemination is an important issue in vehicular networks,
especially considering the vulnerability of vehicle to vehicle (V2V)
message dissemination to malicious attacks. Traditional security mechanisms,
largely based on message encryption and key management, can only guarantee
secure message exchanges between known source and destination pairs.
In vehicular networks however, every vehicle may learn its surrounding
environment and contributes as a source, while in the meantime act
as a destination or a relay of information from other vehicles, message
exchanges often occur between ``stranger'' vehicles. This makes
secure message dissemination against malicious tampering much more
intricate. For secure message dissemination in vehicular networks
against insider attackers, who may tamper the content of the disseminated
messages, ensuring the consistency and integrity of the transmitted
messages becomes a major concern that traditional message encryption
and key management based approaches fall short to provide. However,
it is challenging for a vehicle to distinguish which message is true
when its received messages from multiple nearby vehicles are conflicting.
In this paper, by incorporating the underlying network topology information,
we propose an optimal decision algorithm that is able to maximize
the chance of making a correct decision on the message content, assuming
the prior knowledge of the percentage of malicious vehicles in the
network. Furthermore, a novel heuristic decision algorithm is proposed
that can make decisions without the aforementioned knowledge of the
percentage of malicious vehicles. Simulations are conducted to compare
the security performance achieved by our proposed decision algorithms
with that achieved by existing ones that do not consider or only partially
consider the topological information, to verify the effectiveness
of the algorithms. Our results show that by incorporating the network
topology information, the security performance can be much improved.
This work shed light on the optimum algorithm design for secure message
dissemination. 
\end{abstract}


\section{Introduction}

Vehicular networks, with the assistance of dedicated short-range communication
(DSRC) \cite{Kenny11} and LTE technology, enable safety and non-safety
information sharing among vehicles and infrastructure through vehicle
to vehicle (V2V) and vehicle to infrastructure (V2I) communications,
and therefore are conductive to improving road safety, enhance traffic
efficiency and increase comfort and convenience to drivers and passengers
\cite{Javed17,Siegel17,Zheng15}. On the other hand, accompanying
these benefits brought along by vehicular network applications is
the urgent security issue that should be addressed. Specifically,
considering the vulnerability of V2V communications, message dissemination
in vehicular networks is susceptible to malicious attacks, e.g., malicious
vehicles who may spread false messages, tamper or drop the received
messages \cite{Fonseca06} to disrupt delivery of authentic messages.
These attacks in vehicular networks could potentially result in catastrophic
consequences like city-wide traffic congestion, traffic crash, even
loss of lives, and therefore are significant security threats to transportation
systems that must be thoroughly investigated before vehicular networks
can be deployed. 

Vehicular network security design should guarantee authentication,
non-repudiation, information integrity, and in some specific application
scenarios, confidentiality, to protect the network against attackers
\cite{Sun10}. Conventional security mechanisms, largely based on
message encryption and key management \cite{Tan16,Petit15}, are effective
to guarantee message integrity against outsider attackers, however
fall short of protecting the integrity of disseminated messages when
there exist insider attackers who possess valid certificates that
can pass the authentication process conducted by the certification
authorities \cite{Yang15,Dietzel14-1}. 

To keep the network message dissemination secure against insider attackers,
the trustworthiness of each vehicle and the integrity of their transmitted
messages are of great importance. Different from traditional security
settings, in vehicular networks, information collection and dissemination
are conducted by distributed vehicles. Quite often, information may
be generated by or received from a vehicle that has never been encountered
before. Moreover, the associated vehicular network topology is constantly
changing considering that both V2V and V2I connections may emerge
opportunistically. These unique characteristics may render the entity-based
trust establishment approach, conducted at each vehicle by monitoring
their instantaneous neighbours' behavior, futile in vehicular networks
because it is challenging to maintain a stable reputation value for
an unknown and fast-moving vehicle. Furthermore, safety-related vehicular
network applications usually require vehicles to respond quickly to
the received messages \cite{Du17}. In such cases, determining the
integrity of the disseminated messages is of greater importance than
the malicious vehicle detection. Therefore, decision algorithms based
on data consistency and integrity check emerge, e.g., \cite{Dietzel13,Raya08,Huang14,Zaidi16,Radak}.
However, when a vehicle receives conflicting messages from different
nearby vehicles, it is not straightforward to assess which message
is true if focusing on data only while ignoring the underlying network
topology information that tells where these messages come from. Indeed,
messages coming from different paths can be correlated when the these
paths share some common nodes. For instance, multiple false messages
may result from the same malicious vehicle shared by multiple paths.
Therefore, taking the underlying topological information into consideration
is essential and beneficial when designing decision algorithms for
vehicles to conduct data consistency check. 

In this paper, we consider vehicular networks containing insider malicious
vehicles that may tamper the content of messages to disrupt their
successful delivery. We are interested in investigating topology-based
decision algorithms to keep vehicles from being misguided by false
messages. To the best of our knowledge, this is the first work that
takes the underlying topology information into consideration when
checking the consistency of messages for secure message dissemination.
Our results shed insight on the optimum decision algorithm design
for vehicular networks to improve security performance. 

The novelty and major contributions of this paper are summarized as
follows: 
\begin{enumerate}
\item By utilizing the underlying network topology information, we propose
two message decision algorithms - the optimum decision algorithm and
a heuristic decision algorithm - to cope with the issue of message
inconsistency caused by insider malicious vehicles in the network,
so as to reduce their impact on the message security performance.
\item The proposed optimum decision algorithm is able to effectively help
a vehicle maximally make a correct decision on the content of the
message, given the topology information and a prior knowledge of the
percentage of malicious vehicles in the network. The proposed heuristic
decision algorithm enables a vehicle to make a decision when receiving
conflicting messages purely based on topology information, without
the need for knowing the percentage of malicious vehicles which can
be difficult to estimate in some circumstances. 
\item Simulation results show that both our proposed algorithms outperform
existing decision algorithms that do not consider or only partially
consider the topological information in terms of secure message dissemination
in vehicular networks. Besides, the proposed heuristic decision algorithm,
which is fairly easy to implement in practice, is sufficient to achieve
a high security performance.
\end{enumerate}
The rest of this paper is organized as follows: Section \ref{sec:Related-Work}
reviews related work. Section \ref{sec:System-Model-and problem formation}
introduces the system model and the problem formation. The optimum
decision algorithm and the heuristic decision algorithm are presented
in Section \ref{sec:Optimum-Decision-Algorithm} and Section \ref{sec:Heuristic-Decision-Algorithm}
respectively. In Section \ref{sec:Simulation-and-Discussion}, we
conduct simulations to validate the effectiveness of our proposed
decision algorithms and discuss their insight. Section \ref{sec:Conclusion-and-Future}
concludes this paper.

\section{Related Work\label{sec:Related-Work}}

For secure message dissemination in vehicular networks against insider
malicious vehicles, the trustworthiness of each vehicle and the integrity
of each transmitted message are two major factors need to be considered.
Accordingly, three misbehavior detection schemes are commonly adopted
to help prevent the disseminated messages from being tampered: entity-centric
misbehavior detection scheme, data-centric misbehavior detection scheme,
and a combined use of both. In the following, we will review works
on these three schemes separately. 

Entity-centric misbehavior detection schemes are commonly conducted
at each vehicle by monitoring their instantaneous neighbors' behavior
to assess their trustworthiness level, so as to filter out malicious
vehicles. In \cite{Gazdar12}, Gazdar \textit{et al.} proposed a dynamic
and distributed trust model based on the use of a Markov chain to
evaluate the evolution of each vehicle's trust value. In \cite{Ahmed17},
Ahmed et al. proposed a trust framework to identify malicious nodes
in the network by evaluating the trust value of each vehicle, and
the trust includes node trust and recommendation trust. In \cite{Haddadou15},
motivated by the job market signaling model, Haddadou \textit{et al.}
proposed a distributed trust model for vehicular ad hoc networks (VANETs)
that is able to gradually detect all malicious nodes as well as boosting
the cooperation of selfish nodes. In \cite{Sedjelmaci15}, Sedjelmaci
\textit{et al.} proposed a lightweight intrusion detection framework
with the help of a clustering algorithm to overcome the challenges
of intermittent and ad hoc monitoring and assessment processes caused
by the high mobility and rapid topology change in vehicular networks. 

Data-centric misbehavior detection schemes focus on the consistency
check of the disseminated data to filter out false data. In \cite{Dietzel13},
Dietzel \textit{et al.} indicated that redundant data forwarding paths
are the most promising technique for effective data consistency check
in a multi-hop information dissemination environment, and proposed
three graph-theoretic metrics to measure the redundancy of dissemination
protocols. In \cite{Raya08}, Raya \textit{et al.} proposed a framework
for vehicular networks to establish data-centric trust, and evaluated
the effectiveness of four data fusion rules. In \cite{Huang14}, Huang
\textit{et al.} firstly demonstrated that information cascading and
oversampling adversely affect the performance of trust management
scheme in VANETs, and then proposed a novel voting scheme that takes
the distance between the transmitter and receiver into account when
assigning weight to the trust level of the received data. In \cite{Zaidi16},
Zaidi \textit{et al.} proposed a rogue node detection system for VANETs
utilizing statistical inference techniques to determine whether the
received data are authentic. In \cite{Radak}, Radak \textit{et al.}
applied a so-called cautious operator to deal with data received from
different sources to detect dangerous events on the road. Their adopted
cautious operator is an extension of the Demper-Shafer theory that
is known to be superior in handling data coming from dependent sources. 

A combined misbehavior detection scheme makes use of both the trust
level of vehicles and the consistency of received data to detect misbehaving
vehicles and filter out incorrect messages. Works adopting the combined
scheme are limited. In \cite{Dhurandher14}, Dhurandher \textit{et
al.} proposed a security algorithm using both node reputation and
data plausibility checks to protect the network against attacks. The
node reputation value is obtained by both direct monitoring and indirect
recommendation from neighbors, to detect misbehaving vehicles; and
the data consistency check is conducted by comparing the received
data with the sensed data by the vehicle's own sensors. In \cite{Li16},
Li \textit{et al.} proposed an attack-resistant trust management scheme
to evaluate the trustworthiness of both data and vehicles in VANETs.
They adopted the Dempster-Shafer theory to combine the data received
from different sources, and then used this combined result to update
the trust value of vehicles for misbehavior detection. 

In summary, all the aforementioned works on protecting vehicular networks
from insider attackers either focused on node trust model establishment
and management to detect misbehaving nodes in the network, or focused
on methods to assess data from different sources to check their consistency,
but did not take the underlying network topological information into
consideration. Our work distinguishes from theirs in that we focus
on the received data itself, and utilize the underlying network topology
information to design the decision algorithms for vehicles to check
data consistency so as to maximally protect the authenticity of the
disseminated messages. 

\section{System Model and Problem Formation\label{sec:System-Model-and problem formation}}

In this section, we first introduce the system model, including the
network model, message dissemination model, and the attack model.
Then, we give a rigorous description of the research problem addressed
in this paper. 

\subsection{Network and Message Dissemination Model\label{subsec:Network-and-Message Dissemination moel}}

We consider a vehicular network where each vehicle has a unique ID
number that is registered in certification authority to represent
its identity, and vehicles cannot forge their own or other vehicles'
ID numbers. 

Specifically, consider that there is a vehicle in the network (termed
as the source vehicle) intending to deliver a message about the road
condition to inform other vehicles further away. The road condition
information can be abnormal situations, e.g., hazardous road conditions
such as traffic accident, slippery road, etc., or normal situation,
e.g., uncongested traffic. We assume that the content of message takes
value from $\{0,1\},$ and $1$ represents abnormal road condition
and $0$ represents normal road condition. It is worth noting that
the road situation can also be described as a multi-variable vector
and these variables can be correlated \cite{Raya08}, e.g., one such
variable can be traffic congestion state and another can be accident
state. We denote the content of message transmitted by the source
vehicle, which represents the actual road condition, by $m_{0}$,
$m_{0}\in\{0,1\}$. Other vehicles do not know the true value of $m_{0}$
a priori. 

The message is forwarded from the source vehicle in a broadcast and
multi-hop \cite{Mao09,Ma15} manner to other vehicles with the help
of relay vehicles. Relay vehicles can be any vehicle along the message
propagation path. Multi-path forwarding makes it challenging for the
attackers to influence all message forwarding paths \cite{Dietzel13},
therefore helps to improve the message security performance of the
network. When a vehicle transmits a message to other vehicles, it
adds its identity information, i.e., ID number, to the message. This
is commonly adopted in vehicular network applications and can be achieved
by some standard signature approach \cite{Javed17,Dietzel14}. Using
this, any vehicle in the network is able to obtain an integrity-protected
path list of its received messages recording the relay vehicles of
each message, and the records cannot be injected and removed by attackers. 

\subsection{Attack Model\label{subsec:Attack-Model}}

We consider insider attackers in this paper. That is, we assume all
vehicles are legitimate vehicles that have passed the authentication
process conducted by the certification authority \cite{Raya08,Zaidi16}.
Vehicles in the network can be classified into two categories: \textit{normal
vehicles,} which behave normally and will forward the received message
without any alteration, and \textit{malicious vehicles, }\textit{\emph{which}}\emph{
}may tamper the received message. Malicious vehicles are uniformly
distributed in the system with proportion $p$. It follows that the
probability of each vehicle being a malicious vehicle is $p$, independent
of the event that another distinct vehicle is a malicious vehicle. 

\begin{figure}[t]
\centering{}\includegraphics[width=7.5cm]{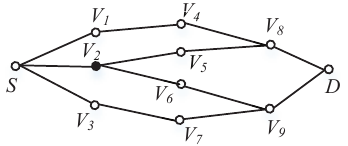} \caption{An illustration of a vehicular network when there exists a malicious
vehicles $V_{2}$ who would tamper the content of message. Specifically,
$S$ is the source vehicle, $D$ is the destination vehicle, and vehicles
between them are relay vehicles. There are four paths ($S-V_{1}-V_{4}-V_{8}-D$,
$S-V_{2}-V_{5}-V_{8}-D$, $S-V_{2}-V_{6}-V_{9}-D$, and $S-V_{3}-V_{7}-V_{9}-D$)
that deliver the source message from $S$ to $D$. Therefore, out
of the four copies of messages received by D, two copies are incorrect
as there are two paths containing the malicious vehicle $V_{2}$.
\label{Fig: System model as an example} }
\end{figure}

Without loss of generality, we assume that the source vehicle is normal
and only relay vehicles may be malicious. The normal vehicles do not
know which vehicles are normal or malicious. On the other hand, malicious
vehicles not only know which vehicles are malicious, but also are
capable of communicating with each other via back channels of infinite
bandwidth \cite{Ponniah17}. That is, we assume that malicious vehicles
know what the correct message transmitted by the source vehicle is.
As a consequence, each malicious vehicle simply transmit the incorrect
message, i.e., different from message $m_{0}$, to its neighbors.
This implies that as long as a message is relayed by at least one
malicious vehicle, the message would be incorrect. Fig. \ref{Fig: System model as an example}
gives a simple example of message dissemination process when there
are insider attackers in the network. 

\subsection{Problem Formation\label{subsec:Problem-formation}}

Now we give a detailed description of the research problem considered
in this paper. 

We consider that there is a vehicle, which is several hops away from
the source vehicle, trying to make a decision on the message content
when it receives several copies of message, and we call it the destination
vehicle. Note that the destination vehicle can be any vehicle along
the message dissemination path. From the time instant the destination
vehicle receives the first message, it waits time period $T$ to receive
more messages before making a final decision. $T$ characterizes the
response time requirement on the decision, and a larger $T$ potentially
allows the vehicle to receive more messages. We will discuss its impact
on the integrity of the decision later in the simulation. Let $k$
be the number of message received by a destination vehicle during
its waiting time period $T$ and let $n$ be the number of relay vehicles
that participate in relaying the $k$ copies of message from the source
vehicle to the destination vehicle. In the following analysis, we
regard $k$ and $n$ are known to the destination vehicle, which can
be readily obtained from the received messages. Consequently, the
network being considered has $n$ relay vehicles and $k$ paths between
the source vehicle and the destination vehicle. Other nodes who do
not participate in the relay have little impact and hence can be ignored.

Denote the $k$ messages received by the destination vehicle by $M_{i}$,
$i=1,2,...k$, $M_{i}\in\{0,1\}$, and let the message vector $M=[M_{1}\;M_{2}\;...M_{k}]^{T}$.
As each message corresponds to a specific delivery path from the source
vehicle to the destination vehicle, we number the corresponding paths
by $L_{1},L_{2},...L_{k}$. In addition, we number the relay vehicles
by $V_{1},V_{2},...V_{n}$. A vehicle $V_{i}$ may belong to one or
more paths.

Note that due to the existence of malicious vehicles who may tamper
the content of the message, the $k$ copies of message received by
the destination vehicle can be in conflict instead of being consistent
with each other. Furthermore, with the potential existence of some
shared relay vehicles in different paths, the $k$ messages received
from $k$ different paths may not be independent. These correlations
are all contained in the information of message dissemination paths.
Therefore, we construct a topology matrix to represent the underlying
network topological correlation. Specifically, based on the path information
derived from the received messages, the destination vehicle can readily
construct a $k\times n$ topology matrix $B$, where each row represents
a path, each column a node (vehicle), and the $(i,j)$-th entry $B_{ij}$
being an indicator whether vehicle $V_{j}$ belongs to path $L_{i}$:
\begin{equation}
B_{ij}=\begin{cases}
1, & \text{if \text{vehicle }\ensuremath{V_{j}} belongs to path\;}L_{i}\\
0, & \text{else}
\end{cases}\label{eq:Definition of topology matrix}
\end{equation}

In this paper, we are interested in designing optimal decision algorithms
for the destination vehicle to maximize the chance of a correct decision
on the content of the disseminated message against attacks from malicious
vehicles by utilizing the underlying network topology information.
Denote by $d$, $d\in\{0,1\}$ the final decision on the content of
message made by the destination vehicle. If the decision is the same
as the source message, i.e., if $d=m_{0}$, we say the destination
vehicle makes a correct decision, otherwise we say it makes an incorrect
decision. We use the probability of correct decision, denoted by $P_{succ}$,
as the performance metric to measure the secure message dissemination
performance, and $P_{succ}$ can be formally defined as follows: 
\begin{equation}
P_{succ}=\text{Pr}(d=1,m_{0}=1)+\text{Pr}(d=0,m_{0}=0)\label{eq:definition of Psucc}
\end{equation}

\section{Optimum Decision Algorithm\label{sec:Optimum-Decision-Algorithm}}

In this section, we propose a decision algorithm aims to optimize
the secure message dissemination performance in terms of maximizing
the probability of correct decision $P_{succ}$, that is, 
\begin{align}
\max & \;P_{succ},\label{eq:formation of optimization problem_maximize Psucc}
\end{align}
where $P_{succ}$ is given by \eqref{eq:definition of Psucc}. 

In the following, we will first present the optimum decision algorithm
followed by a detailed proof to prove its optimality, and then we
will introduce its detailed implementation and discuss its limitation
in practical realization.

\subsection{Optimum Decision Algorithm\label{subsec:Optimum-Decision-Algorithm description}}

The following theorem summarizes the optimum decision algorithm to
maximize $P_{succ}$.
\begin{thm}
\label{thm:Optimum decision algorithm}Consider that a destination
vehicle receives $k$ copies of messages $M_{1}=m_{1},M_{2}=m_{2},...M_{k}=m_{k}$.
Given the prior knowledge of the probabilities that the occurrence
of abnormal event of interest, e.g., traffic congestion, are $P_{1}=\text{Pr}(m_{0}=1)$,
and $P_{0}=1-P_{1}=\text{Pr}(m_{0}=0)$, which can be estimated from
empirical knowledge \cite{Wang03}, the optimum decision algorithm
that leads to \eqref{eq:formation of optimization problem_maximize Psucc}
can be shown as follows:
\begin{equation}
d=\begin{cases}
1, & \frac{\text{Pr}\left(M_{1}=m_{1},...M_{k}=m_{k}|m_{0}=1\right)}{\text{Pr}\left(M_{1}=m_{1},...M_{k}=m_{k}|m_{0}=0\right)}>\frac{P_{0}}{P_{1}}\\
0, & \frac{\text{Pr}\left(M_{1}=m_{1},...M_{k}=m_{k}|m_{0}=1\right)}{\text{Pr}\left(M_{1}=m_{1},...M_{k}=m_{k}|m_{0}=0\right)}<\frac{P_{0}}{P_{1}}
\end{cases},\label{eq:optimal fusion rule to maximize Psucc}
\end{equation}
and when $\frac{\text{Pr}\left(M_{1}=m_{1},...M_{k}=m_{k}|m_{0}=1\right)}{\text{Pr}\left(M_{1}=m_{1},...M_{k}=m_{k}|m_{0}=0\right)}=\frac{P_{0}}{P_{1}}$,
$d$ is randomly chosen from 0 and 1 with equal probability. 
\end{thm}
\begin{IEEEproof}
As introduced in \cite{Kam92,Zhu02}, the objective of a binary Bayes
decision problem is to minimize the expectation of the decision cost,
denoted by $U(d,m_{0})$. Let $U_{ij}$, $i=0,1$, $j=0,1$, represents
the cost of declaring the final result $d=i$ when actually the source
message $m_{0}=j\neq i$, and $U_{ij}$ can be negative to represent
the benefits of making a correct decision. As a ready consequence
of the total probability theorem, the expectation of the decision
cost $U(d,m_{0})$ can be expressed as follows: 
\begin{equation}
U(d,m_{0})=\sum_{i=0}^{1}\sum_{j=0}^{1}U_{ij}\text{Pr}\left(d=i,\;m_{0}=j\right).\label{eq:definition of Bayes cost}
\end{equation}

When assuming $U_{01}>U_{11}$ and $U_{10}>U_{00}$, which is reasonable
considering the cost of making an incorrect decision is usually larger
than that making a correct decision, the optimum decision algorithm
that minimizes the expectation of the decision cost made by the destination
vehicle given its $k$ copies of received message $M_{1}=m_{1},M_{2}=m_{2},...M_{k}=m_{k}$,
is given by \cite{Zhu02}:
\begin{equation}
d=\begin{cases}
1 & \frac{\text{Pr}\left(M_{1}=m_{1},...M_{k}=m_{k}|m_{0}=1\right)}{\text{Pr}\left(M_{1}=m_{1},...M_{k}=m_{k}|m_{0}=0\right)}>\frac{P_{0}\left(U_{10}-U_{00}\right)}{P_{1}\left(U_{01}-U_{11}\right)}\\
0, & \frac{\text{Pr}\left(M_{1}=m_{1},...M_{k}=m_{k}|m_{0}=1\right)}{\text{Pr}\left(M_{1}=m_{1},...M_{k}=m_{k}|m_{0}=0\right)}<\frac{P_{0}\left(U_{10}-U_{00}\right)}{P_{1}\left(U_{01}-U_{11}\right)}
\end{cases},\label{eq:fusion rule of bayes decision problem}
\end{equation}
where $\text{Pr}\left(M_{1}=m_{1},...M_{k}=m_{k}|m_{0}=1\right)$
and $\text{Pr}\left(M_{1}=m_{1},...M_{k}=m_{k}|m_{0}=0\right)$ are
the two conditional probabilities of the occurrence of event $M_{1}=m_{1},M_{2}=m_{2},...M_{k}=m_{k}$,
which characterize the correlations between received messages. Besides,
when a tie occurs, namely, when $\frac{\text{Pr}\left(M_{1}=m_{1},...M_{k}=m_{k}|m_{0}=1\right)}{\text{Pr}\left(M_{1}=m_{1},...M_{k}=m_{k}|m_{0}=0\right)}=\frac{P_{0}\left(U_{10}-U_{00}\right)}{P_{1}\left(U_{01}-U_{11}\right)}$,
$d$ is randomly chosen from $0$ and 1 with equal probability. 

From \eqref{eq:definition of Bayes cost}, when assuming the cost
of making a correct decision is $0$ and making an incorrect decision
is $1$, namely, by assuming $U_{00}=U_{11}=0$ and $U_{01}=U_{10}=1$,
we have:
\begin{align}
U(d,m_{0}) & =\text{Pr}\left(d=0,m_{0}=1\right)+\text{Pr}\left(d=1,m_{0}=0\right)\nonumber \\
 & =1-P_{succ}.\label{eq:special case of cost}
\end{align}
It follows that a minimization of the expectation of the decision
cost, is equivalent to a maximization of the probability of correct
decision, namely, we have
\begin{equation}
\min\;U(d,m)\;\Longleftrightarrow\max\;P_{succ}\label{eq:Problem transformation}
\end{equation}
Therefore, the optimum decision algorithm for the optimization problem
\eqref{eq:formation of optimization problem_maximize Psucc} is exactly
the decision algorithm that provides a solution to the classical Bayes
decision problem in a special case, shown as \eqref{eq:optimal fusion rule to maximize Psucc},
which finalize the proof. 
\end{IEEEproof}
\begin{rem}
\label{rem:Discuss the Opt-algo}It can be seen from \eqref{eq:optimal fusion rule to maximize Psucc}
that, given the probabilities of the occurrence of abnormal event
of interest, $P_{0}$ and $P_{1}$ respectively, the decision on $d=1$
or $d=0$ depends on the ratio $\frac{\text{Pr}\left(M_{1}=m_{1},...M_{k}=m_{k}|m_{0}=1\right)}{\text{Pr}\left(M_{1}=m_{1},...M_{k}=m_{k}|m_{0}=0\right)}$.
That is, given a set of received messages $M_{1}=m_{1},M_{2}=m_{2},...M_{k}=m_{k}$,
the destination vehicle need to calculate the probability that the
event $M_{1}=m_{1},...M_{k}=m_{k}$ occurs if the true message $m_{0}$
is 1, denoted as $\text{Pr}\left(M_{1}=m_{1},...M_{k}=m_{k}|m_{0}=1\right)$,
and the probability that the event occurs if the true message $m_{0}$
is 0, denoted as $\text{Pr}\left(M_{1}=m_{1},...M_{k}=m_{k}|m_{0}=0\right)$.
A decision on $d$ is then made by comparing the value of $\frac{\text{Pr}\left(M_{1}=m_{1},...M_{k}=m_{k}|m_{0}=1\right)}{\text{Pr}\left(M_{1}=m_{1},...M_{k}=m_{k}|m_{0}=0\right)}$
and $\frac{P_{0}}{P_{1}}$. Therefore, calculation of the two probabilities
is the critical part of implementing the algorithm in practice. 
\end{rem}
In summary, the optimum decision algorithm for the destination vehicle
to maximally make a correct decision on the message content works
as detailed in Algorithm \ref{alg:Optimum-decision-algorithm}, where
the details of calculating the two terms $\text{Pr}\left(M_{1}=m_{1},...M_{k}=m_{k}|m_{0}=1\right)$
and $\text{Pr}\left(M_{1}=m_{1},...M_{k}=m_{k}|m_{0}=0\right)$ will
be given in the following subsection. 

\begin{algorithm}[h]
\caption{Optimum decision algorithm\label{alg:Optimum-decision-algorithm}}
\begin{lyxlist}{00.00.0000}
\item [{\textbf{INPUT}:}] $M_{1},M_{2},...M_{k}$, $P_{0}$, $P_{1}$,
$p$
\item [{\textbf{OUTPUT}:}] $d$
\item [{\textbf{begin}}]~
\end{lyxlist}
\begin{enumerate}
\item Construct topology matrix $B$ based on the paths information derived
from the received $k$ copies of message;
\item Calculate $\text{Pr}\left(M_{1}=m_{1},...M_{k}=m_{k}|m_{0}=1\right)$
and $\text{Pr}\left(M_{1}=m_{1},...M_{k}=m_{k}|m_{0}=0\right)$ according
to \eqref{eq:Calculation of conditional joint pmf conditioned on m0=00003D1, n0>0}
and \eqref{eq:Calculation of conditional joint pmf conditioned on m0=00003D0, n1>0}
respectively, given the network topology information and a prior knowledge
on the proportion of malicious vehicles in the network;
\item \textbf{If} $\frac{\text{Pr}\left(M_{1}=m_{1},...M_{k}=m_{k}|m_{0}=1\right)}{\text{Pr}\left(M_{1}=m_{1},...M_{k}=m_{k}|m_{0}=0\right)}>\frac{P_{0}}{P_{1}}$
\textbf{then} $d=1;$\\
\\
\textbf{elseif} $\frac{\text{Pr}\left(M_{1}=m_{1},...M_{k}=m_{k}|m_{0}=1\right)}{\text{Pr}\left(M_{1}=m_{1},...M_{k}=m_{k}|m_{0}=0\right)}<\frac{P_{0}}{P_{1}}$
\textbf{then} $d=0$;\\
\\
\textbf{else then $d$ }is randomly chosen from $0$ and $1$ with
equal probability\textbf{;}\\
\textbf{}\\
\textbf{end}
\end{enumerate}
\begin{lyxlist}{00.00.0000}
\item [{\textbf{end}}]~
\end{lyxlist}
\end{algorithm}

\subsection{Algorithm Implementation\label{subsec:Optimum-Algorithm-Implementation}}

In this part, we will introduce the detailed implementation of the
proposed optimum decision algorithm. As discussed in Remark \ref{rem:Discuss the Opt-algo},
the first step is to calculate the two probabilities $\text{Pr}\left(M_{1}=m_{1},...M_{k}=m_{k}|m_{0}=1\right)$
and $\text{Pr}\left(M_{1}=m_{1},...M_{k}=m_{k}|m_{0}=0\right)$ as
they are prerequisite to obtaining the final decision $d$. 

The main idea behind the calculation of $\text{Pr}\left(M_{1}=m_{1},...M_{k}=m_{k}|m_{0}=1\right)$
and $\text{Pr}\left(M_{1}=m_{1},...M_{k}=m_{k}|m_{0}=0\right)$ is
as follows. We classify vehicles into three different types based
on the paths they belong to. We call a vehicle a \textit{Type 0} (or
\textit{Type 1}) vehicle if it only belongs to paths that deliver
messages with content 0 (or 1) to the destination vehicle, and a vehicle
is a \textit{Type 2} vehicle (if any) if it belongs to at least one
path that delivers message with content 0 and another path that delivers
message with content 1 to the destination vehicle. That is a Type
0 vehicle only belongs to paths that deliver consistent messages 0;
a Type 1 vehicle only belongs to paths that deliver consistent messages
1; while a Type 2 vehicle belong to paths that delivers inconsistent
messages. Therefore, by separating the paths according to the delivered
message contents, the conclusion readily follows that given $m_{0}=1$,
all the Type 1 and Type 2 vehicles are normal vehicles, meanwhile
malicious vehicles only exist among Type 0 vehicles. Then, by listing
and analyzing all the different combination of malicious vehicles
among the Type 0 vehicles, we can obtain the result of our target
conditional probability $\text{Pr}\left(M_{1}=m_{1},...M_{k}=m_{k}|m_{0}=1\right)$.
The idea of calculating $\text{Pr}\left(M_{1}=m_{1},...M_{k}=m_{k}|m_{0}=0\right)$
is totally the same. 

In the following, we will first demonstrate the method of constructing
topology matrix $B$ based on the above idea, and then calculate the
two probabilities $\text{Pr}\left(M_{1}=m_{1},...M_{k}=m_{k}|m_{0}=1\right)$
and $\text{Pr}\left(M_{1}=m_{1},...M_{k}=m_{k}|m_{0}=0\right)$ respectively.
Without loss of generality, we assume that among the $k$ copies of
messages $M_{1}=m_{1},...M_{k}=m_{k}$ received by the destination
vehicle, there are exactly $k_{1}$, messages with content $1$ and
the other $k-k_{1}$ messages with content $0$. Note that $k_{1}=0$
and $k_{1}=k$ are both trivial cases implying no conflict in the
received messages so that the decision is straightforward, therefore
we only consider the case when $0<k_{1}<k$. 

\subsubsection{Constructing the topology matrix $B$}

Specifically, recall the definition of the topology matrix given in
\eqref{eq:Definition of topology matrix}, that each row corresponds
to a path and each column corresponds to a vehicle. Based on the idea
discussed above to calculate the probabilities $\text{Pr}\left(M_{1}=m_{1},...M_{k}=m_{k}|m_{0}=1\right)$
and $\text{Pr}\left(M_{1}=m_{1},...M_{k}=m_{k}|m_{0}=0\right)$, we
re-arrange the network topology matrix $B$ in the following form:
\begin{equation}
B=\left[\begin{array}{ccc}
B_{1} & B_{s_{1}} & \mathbf{\boldsymbol{0}}\\
\mathbf{\boldsymbol{0}} & B_{s_{0}} & B_{0}
\end{array}\right],
\end{equation}
where $B_{1}$, $B_{0}$, $B_{s_{1}}$ and $B_{s_{0}}$, if exist,
are non-zero matrices, and $\left[\begin{array}{ccc}
B_{1} & B_{s_{1}} & \mathbf{\boldsymbol{0}}\end{array}\right]$ is a $k_{1}\times n$ sub-matrix corresponding to the paths that
deliver messages with content 1 to the destination vehicle, and $\left[\begin{array}{ccc}
\mathbf{\boldsymbol{0}} & B_{s_{0}} & B_{0}\end{array}\right]$ is a $(k-k_{1})\times n$ sub-matrix corresponding to the paths that
deliver messages with content 0 to the destination vehicle. Besides,
the columns of $B_{1}$ and $B_{0}$ correspond to vehicles that only
belong to paths that deliver messages with content 1 and that deliver
messages with content 0 to the destination vehicle respectively, i.e.,
Type 1 vehicles and Type 0 vehicles respectively. The columns of sub-matrix
$\left[\begin{array}{c}
B_{s_{1}}\\
B_{s_{0}}
\end{array}\right]$ correspond to all the Type 2 vehicles. Assume that the number of
Type $1$ and Type $0$ vehicles are $n_{1}$ and $n_{0}$ respectively,
$0\leq n_{1}+n_{0}\leq n$, and the number of Type 2 vehicles is $n_{2}=n-n_{1}-n_{0}$.
It follows that matrices $B_{1}$ and $B_{0}$ are of size $k_{1}\times n_{1}$
and $(k-k_{1})\times n_{0}$ respectively, and the matrix $\left[\begin{array}{c}
B_{s_{1}}\\
B_{s_{0}}
\end{array}\right]$ is of size $k\times\left(n-n_{1}-n_{0}\right)$. 

It is worth noting that the above arrangement of columns and rows
of matrix $B$ corresponds to a re-numbering of vehicles and paths
and it does not change the underlying topology in terms of paths information.
Besides, the sub-matrix $B_{1}$ can be non-existent if $n_{1}=0$,
i.e., when the paths that deliver messages 0 to the destination vehicle
contains all the $n$ vehicles in the network. Under this circumstance,
$B=\left[\begin{array}{cc}
B_{s_{1}} & \mathbf{\boldsymbol{0}}\\
B_{s_{0}} & B_{0}
\end{array}\right]$. Similarly, the sub-matrix $B_{0}$ (or $\left[\begin{array}{c}
B_{s_{1}}\\
B_{s_{0}}
\end{array}\right]$) can also be non-existent when $n_{0}=0$ (or $n_{2}=0)$. 

\subsubsection{Calculation of $\text{Pr}\left(M_{1}=m_{1},...M_{k}=m_{k}|m_{0}=1\right)$
and $\text{Pr}\left(M_{1}=m_{1},...M_{k}=m_{k}|m_{0}=0\right)$ }

In this part, we show the method of calculating the two conditional
probabilities $\text{Pr}\left(M_{1}=m_{1},...M_{k}=m_{k}|m_{0}=1\right)$
and $\text{Pr}\left(M_{1}=m_{1},...M_{k}=m_{k}|m_{0}=0\right)$ using
the constructed topology matrix $B$. The following two theorems summarize
the results.
\begin{thm}
\label{thm:calculationf of Pr(M1,...Mk|m0=00003D1)}Consider that
a destination vehicle receives $k$ copies of message $M_{1}=m_{1},M_{2}=m_{2},...M_{k}=m_{k}$,
and among which $k_{1}$ messages are with content $1$ and the other
$k-k_{1}$ messages are with content 0, $0<k_{1}<k$. Conditioned
on the source message $m_{0}=1$, the conditional probability of the
occurrence of event $M_{1}=m_{1},...M_{k}=m_{k}$ can be calculated
as follows: 
\begin{align}
 & \text{Pr}\left(M_{1}=m_{1},...M_{k}=m_{k}|m_{0}=1\right)\nonumber \\
= & \begin{cases}
\left(1-p\right)^{n-n_{0}}\cdot\left[\sum_{i=1}^{n_{0}}a_{i}\cdot p^{i}\left(1-p\right)^{n_{0}-i}\right], & n_{0}>0\\
0, & n_{0}=0
\end{cases},\label{eq:Calculation of conditional joint pmf conditioned on m0=00003D1}
\end{align}
where $n_{0}$ is the number of vehicles that only belong to paths
that deliver messages with content 0 to the destination vehicle, i.e.,
the number of Type 0 vehicles in the network, and $a_{i},i=1,2,...n_{0}$
is the number of combinations that contain exactly $i$ malicious
Type 0 vehicles leading to the occurrence of event $M_{1}=m_{1},...M_{k}=m_{k}$. 
\end{thm}
\begin{IEEEproof}
When $n_{0}=0$, there are no Type 0 vehicles in the network, which
implies that the paths that deliver messages with content $1$ to
the destination vehicle contain all the $n$ vehicles in the network,
and the topology matrix $B=\left[\begin{array}{cc}
B_{1} & B_{s_{1}}\\
\mathbf{\boldsymbol{0}} & B_{s_{0}}
\end{array}\right]$. Under this circumstance, conditioned on the source message $m_{0}=1$,
when the event that $k_{1}$ messages are with content 1 occurs, all
the $n$ vehicles in the network should be normal vehicles. It follows
that the event that the other $k-k_{1}$ messages are with content
0 occurs with probability 0. Therefore, we have $\text{Pr}\left(M_{1}=m_{1},...M_{k}=m_{k}|m_{0}=1\right)=0$
when $n_{0}=0$. 

When $n_{0}>0$, from the topology matrix $B$, we can conclude that
if the matrix $\left[\begin{array}{c}
B_{s_{1}}\\
B_{s_{0}}
\end{array}\right]$ exists, then the corresponding Type 2 vehicles should be all normal
vehicles. Observing that there is no possibility for two paths sharing
the same malicious vehicle to deliver different contents. Therefore,
malicious vehicles exist either among Type 1 vehicles or among Type
0 vehicles. 

Given the source message $m_{0}=1$, all the Type 1 vehicles should
be normal vehicles. Malicious vehicles can only exist among Type 0
vehicles. Besides, the malicious Type 0 vehicles should be able to
compromise all the $k-k_{1}$ paths (corresponding to the sub-matrix
$\left[\begin{array}{ccc}
\mathbf{\boldsymbol{0}} & B_{s_{0}} & B_{0}\end{array}\right]$) to cause the occurrence of the event that all the $k-k_{1}$ paths
delivering messages with incorrect content $0$. Therefore, any combination
of malicious vehicles should satisfy the follows condition: by implementing
element-wise \textit{union} on their corresponding columns in sub-matrix
$B_{0}$, i.e., implementing element-wise Boolean operation \textit{OR}
on them, the result should be a column with each entry be $1$. 

Note that the number of malicious type 0 vehicles can be any integer
within $[1,n_{0}]$. We denote by event $e_{i}$ that randomly choosing
$i$ columns from sub-matrix $B_{0}$ and then conducting element-wise
union operation to them, there results a column with each entry being
$1$. Denote by $a_{i},i=1,2,...n_{0}$ the total number of combinations
that event $e_{i}$ occurs. Therefore, we have 
\begin{equation}
a_{i}=\sum_{j=1}^{z_{i}}I\left(\text{event}\;e_{i}\;\text{occurs}\right),\label{definiton of a_i}
\end{equation}
where $z_{i}=\left(\begin{array}{c}
n_{0}\\
i
\end{array}\right)$, and $I(x)$ is an indicator function that $I(x)=1,$ when $x$ is
true; otherwise $I(x)=0$. 

It then follows from the combination theory \cite{Feller71} that
:
\begin{align}
 & \text{Pr}\left(M_{1}=m_{1},...M_{k}=m_{k}|m_{0}=1\right)\nonumber \\
= & \left(1-p\right)^{n-n_{0}}\cdot\left[\sum_{i=1}^{n_{0}}a_{i}\cdot p^{i}\left(1-p\right)^{n_{0}-i}\right],\label{eq:Calculation of conditional joint pmf conditioned on m0=00003D1, n0>0}
\end{align}
where the first part corresponds to the probability that the $k_{1}$
paths deliver messages with correct content $1$, so that all the
$n-n_{0}$ vehicles contained in these $k_{1}$ paths are therefore
normal vehicles; and the second part is the probability that the $k-k_{1}$
paths deliver messages with incorrect content $0$, which summing
up all the probabilities of different malicious vehicle combinations. 
\end{IEEEproof}
\begin{thm}
\label{thm:calculation of Pr(M1,...Mk|m0=00003D0)}Consider that a
destination vehicle receives $k$ copies of message $M_{1}=m_{1},M_{2}=m_{2},...M_{k}=m_{k}$,
and among which $k_{1}$ messages are with content $1$ and the other
$k-k_{1}$ messages are with content 0, $0<k_{1}<k$. Conditioned
on the source message $m_{0}=0$, the conditional probability of the
occurrence of event $M_{1}=m_{1},...M_{k}=m_{k}$ can be calculated
as follows: 
\begin{align}
 & \text{Pr}\left(M_{1}=m_{1},...M_{k}=m_{k}|m_{0}=0\right)\nonumber \\
= & \begin{cases}
\left(1-p\right)^{n-n_{1}}\cdot\left[\sum_{i=1}^{n_{1}}b_{i}\cdot p^{i}\left(1-p\right)^{n_{1}-i}\right], & n_{1}>0\\
0, & n_{1}=0
\end{cases},\label{eq:Calculation of conditional joint pmf conditioned on m0=00003D0}
\end{align}
where $n_{1}$ is the number of vehicles that only belong to paths
that deliver messages with content 1 to the destination vehicle, i.e.,
the number of Type 1 vehicles in the network, and $b_{i},i=1,2,...n_{1}$
is the number of combinations that exactly $i$ malicious Type 1 vehicles
leading to the occurrence of event $M_{1}=m_{1},...M_{k}=m_{k}$. 
\end{thm}
Denote by event $e_{i}^{'}$ that randomly choosing $i$ columns from
sub-matrix $B_{1}$ and then conducting element-wise union operation
to them, there results a column with each entry be $1$. Denote by
$b_{i},i=1,2,...n_{1}$ the total number of combinations that event
$e_{i}^{'}$ occurs. Then we have 
\begin{equation}
b_{i}=\sum_{j=1}^{z_{i}^{'}}I\left(\text{event}\;e_{i}^{'}\;\text{occurs}\right),\label{definiton of b_i}
\end{equation}
where $z_{i}^{'}=\left(\begin{array}{c}
n_{1}\\
i
\end{array}\right)$. Therefore, this theorem can be readily proved following the same
method as that used in the proof of Theorem \ref{thm:calculationf of Pr(M1,...Mk|m0=00003D1)},
and hence is ignored. 

\subsection{Discussion\label{subsec:Discussion of Optimum Algorithm}}

From the analysis in Section \ref{subsec:Optimum-Algorithm-Implementation},
we can see that the value of $n_{0}$, $n_{1}$, and $a_{i}$, $i=1,2,...n_{0}$
in \eqref{eq:Calculation of conditional joint pmf conditioned on m0=00003D1},
$b_{i}$, $i=1,2,...n_{1}$ in \eqref{eq:Calculation of conditional joint pmf conditioned on m0=00003D0}
can be obtained from the network topology matrix. That is, when the
$k$ received messages $M_{1}=m_{1},...M_{k}=m_{k}$, and the network
topology is given, the value of $n_{0}$, $n_{1}$, $a_{i}$, $i=1,2,...n_{0}$,
and $b_{i}$, $i=1,2,...n_{1}$ are all determined. However, the exact
values of $\text{Pr}\left(M_{1}=m_{1},...M_{k}=m_{k}|m_{0}=1\right)$
and $\text{Pr}\left(M_{1}=m_{1},...M_{k}=m_{k}|m_{0}=0\right)$, shown
also in \eqref{eq:Calculation of conditional joint pmf conditioned on m0=00003D1}
and \eqref{eq:Calculation of conditional joint pmf conditioned on m0=00003D0},
also depend on the proportion of malicious vehicles $p$ in the network,
which usually, is not easy to be obtained or estimated as a prior
knowledge. In the following, we use a simple example to show the dependency
on $p$ of the proposed optimum decision algorithm. 

\begin{figure}[t]
\centering{}\includegraphics[width=8cm]{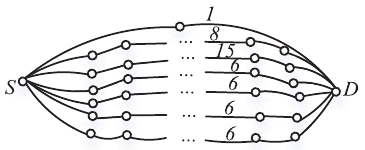}\caption{An illustration of a vehicular network that contains $7$ independent
paths from the source vehicle to the destination vehicle, each path
containing 1, 8, 15, 6, 6, 6, 6 vehicles respectively. \label{Fig: An example to illustrate the dependence on p} }
\end{figure}

Consider a network that contains a total of $7$ independent paths
from the source vehicle to the destination vehicle. The first three
paths, containing $1,$ $8$ and $15$ vehicles respectively deliver
messages with content $1$ to the destination vehicle, and the other
four paths, containing $6$ vehicles each, deliver messages with content
$0$ to the destination vehicle. See Fig. \ref{Fig: An example to illustrate the dependence on p}
for an illustration. 

\begin{figure}[t]
\centering{}\includegraphics[width=7cm]{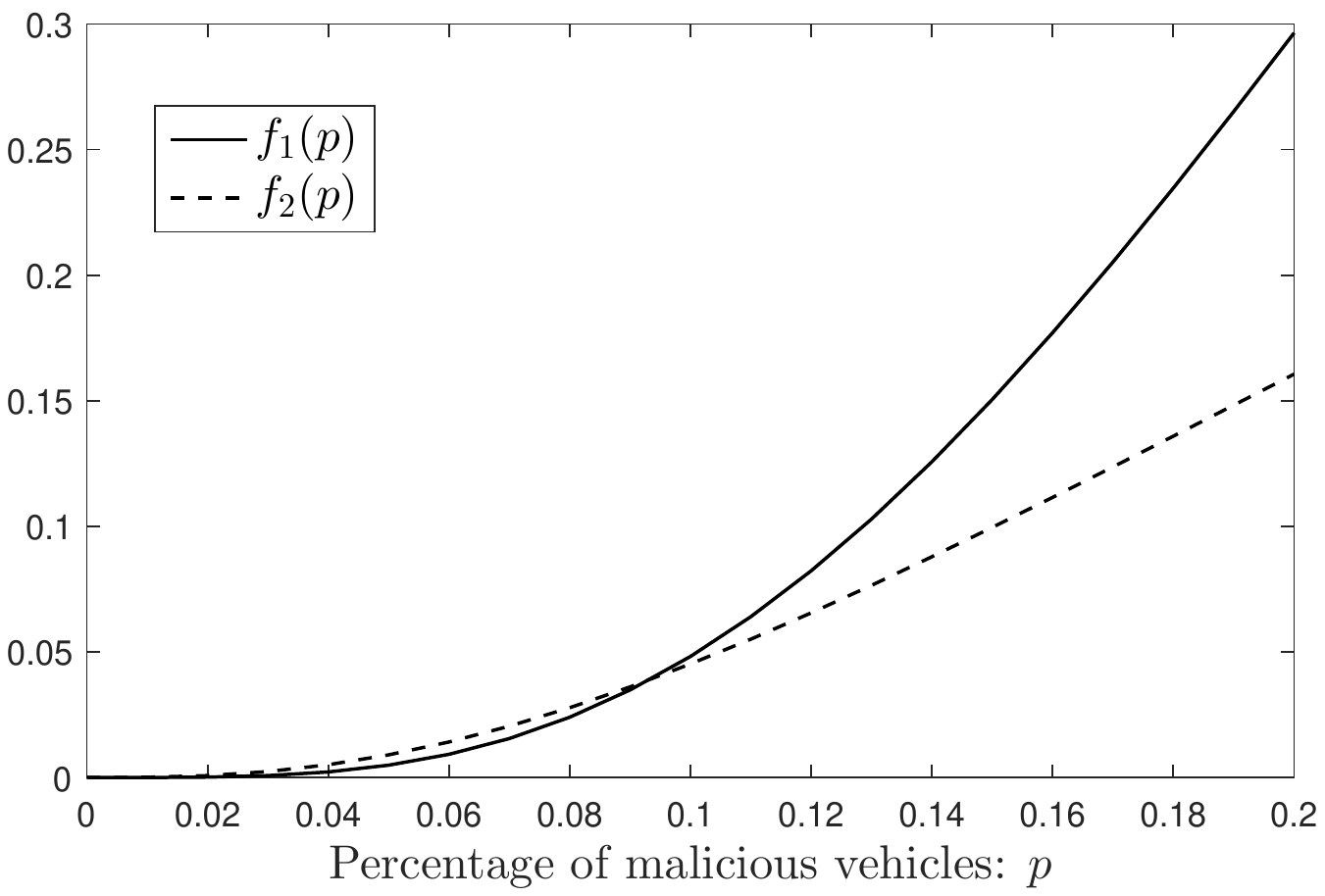}\caption{An illustration to show that the percentage of malicious vehicles
is indispensable in implementing the optimum decision algorithm to
achieve an accurate decision result. \label{Fig: illustration of dependence on p} }
\end{figure}

According to \eqref{eq:Calculation of conditional joint pmf conditioned on m0=00003D1}
and \eqref{eq:Calculation of conditional joint pmf conditioned on m0=00003D0},
we have:
\begin{align}
 & \text{Pr}\left(M_{1}=M_{2}=M_{3}=1,M_{4}=...=M_{7}=0|m_{0}=1\right)\nonumber \\
= & (1-p)^{1+8+15}\cdot\left[1-(1-p)^{6}\right]^{4}\nonumber \\
= & (1-p)^{24}\cdot\left[1-(1-p)^{6}\right]^{4},
\end{align}
and
\begin{align}
 & \text{Pr}\left(M_{1}=M_{2}=M_{3}=1,M_{4}=...=M_{7}=0|m_{0}=0\right)\nonumber \\
= & (1-p)^{6\times4}\cdot\left[1-\left(1-p\right)\right]\left[1-(1-p)^{8}\right]\left[1-(1-p)^{15}\right]\nonumber \\
= & (1-p)^{24}p\left[1-(1-p)^{8}\right]\left[1-(1-p)^{15}\right].
\end{align}
Therefore, 
\begin{align}
 & \frac{\text{Pr}\left(M_{1}=M_{2}=M_{3}=1,M_{4}=...=M_{7}=0|m_{0}=1\right)}{\text{Pr}\left(M_{1}=M_{2}=M_{3}=1,M_{4}=...=M_{7}=0|m_{0}=0\right)}\nonumber \\
= & \frac{1-(1-p)^{6}}{p\left[1-(1-p)^{8}\right]\left[1-(1-p)^{15}\right]}
\end{align}
Let 
\begin{equation}
f_{1}(p)=1-(1-p)^{6}
\end{equation}
and 
\begin{equation}
f_{2}(p)=p\left[1-(1-p)^{8}\right]\left[1-(1-p)^{15}\right],
\end{equation}
and plot them with different values of $p$, see Fig. \ref{Fig: illustration of dependence on p}
for an illustration. We can see that the value of $\frac{\text{Pr}\left(M_{1}=m_{1},...M_{k}=m_{k}|m_{0}=1\right)}{\text{Pr}\left(M_{1}=m_{1},...M_{k}=m_{k}|m_{0}=0\right)}$
depends on the percentage of malicious vehicles in the network. Specifically,
it is shown in Fig. \ref{Fig: illustration of dependence on p} that
when $p$ is smaller than a threshold, e.g., $p_{th}=0.092$ in this
case, the value of $\frac{\text{Pr}\left(M_{1}=m_{1},...M_{k}=m_{k}|m_{0}=1\right)}{\text{Pr}\left(M_{1}=m_{1},...M_{k}=m_{k}|m_{0}=0\right)}=\frac{f_{1}(p)}{f_{2}(p)}$
is smaller than $1$, while when $p$ is larger than the threshold,
the value of $\frac{\text{Pr}\left(M_{1}=m_{1},...M_{k}=m_{k}|m_{0}=1\right)}{\text{Pr}\left(M_{1}=m_{1},...M_{k}=m_{k}|m_{0}=0\right)}=\frac{f_{1}(p)}{f_{2}(p)}$
is larger than $1$, and will further increase with an increase of
$p$. Therefore, given the network topology, the optimum decision
based on \eqref{eq:optimal fusion rule to maximize Psucc} relies
on the value of $p$. This illustrates that the value of $p$ is indispensable
in adopting the optimum decision algorithm to achieve an accurate
decision result. 

\section{Heuristic Decision Algorithm\label{sec:Heuristic-Decision-Algorithm}}

As discussed in the Section \ref{subsec:Discussion of Optimum Algorithm},
the implementation of the optimum decision algorithm proposed in the
last section relies on prior knowledge of the percentage of malicious
vehicles $p$ in the network, which is usually not easy to be obtained
or estimated. In this section, to eliminate the dependence on $p$,
we propose a heuristic decision algorithm for the destination vehicle
to make a decision when receiving conflicting messages purely based
on network topology information only.

The heuristic decision algorithm is derived from the principle of
Maximum Likelihood Estimation \cite{DeGroot02}, which can be described
as follows:

\begin{equation}
d=\begin{cases}
1, & \frac{\text{Pr}\left(M_{1}=m_{1},...M_{k}=m_{k}|m_{0}=1\right)}{\text{Pr}\left(M_{1}=m_{1},...M_{k}=m_{k}|m_{0}=0\right)}>1\\
0, & \frac{\text{Pr}\left(M_{1}=m_{1},...M_{k}=m_{k}|m_{0}=1\right)}{\text{Pr}\left(M_{1}=m_{1},...M_{k}=m_{k}|m_{0}=0\right)}<1
\end{cases},\label{eq:MLE rule}
\end{equation}
where $M_{1}=m_{1},...M_{k}=m_{k}$ are the $k$ messages received
by the destination vehicle, $m_{0}$ is the source message and $d$
is the decision made by the destination vehicle. When $\frac{\text{Pr}\left(M_{1}=m_{1},...M_{k}=m_{k}|m_{0}=1\right)}{\text{Pr}\left(M_{1}=m_{1},...M_{k}=m_{k}|m_{0}=0\right)}=1$,
$d$ is randomly chosen from 0 and 1 with equal probability. 

Based on the received messages $M_{1}=m_{1}$, $M_{2}=m_{2}$, $...$,
$M_{k}=m_{k}$ and the path information obtained from messages, the
method of constructing the topology matrix $B$ is the same as introduced
in Section \ref{subsec:Optimum-Algorithm-Implementation}, i.e., $B=\left[\begin{array}{ccc}
B_{1} & B_{s_{1}} & \mathbf{\boldsymbol{0}}\\
\mathbf{\boldsymbol{0}} & B_{s_{0}} & B_{0}
\end{array}\right]$. Therefore, by combining \eqref{eq:Calculation of conditional joint pmf conditioned on m0=00003D1},
\eqref{eq:Calculation of conditional joint pmf conditioned on m0=00003D0}
and \eqref{eq:MLE rule}, it is ready to have $d=\begin{cases}
0, & n_{0}=0\\
1, & n_{1}=0
\end{cases}$, and when $n_{0}>0$ and $n_{1}>0$,
\begin{align}
 & \frac{\text{Pr}\left(M_{1}=m_{1},...M_{k}=m_{k}|m_{0}=1\right)}{\text{Pr}\left(M_{1}=m_{1},...M_{k}=m_{k}|m_{0}=0\right)}\nonumber \\
= & \frac{\left(1-p\right)^{n-n_{0}}\cdot\left[\sum_{i=1}^{n_{0}}a_{i}\cdot p^{i}\left(1-p\right)^{n_{0}-i}\right]}{\left(1-p\right)^{n-n_{1}}\cdot\left[\sum_{i=1}^{n_{1}}b_{i}\cdot p^{i}\left(1-p\right)^{n_{1}-i}\right]}\nonumber \\
= & \frac{\sum_{i=1}^{n_{0}}a_{i}\cdot\left(\frac{p}{1-p}\right){}^{i}}{\sum_{i=1}^{n_{1}}b_{i}\cdot\left(\frac{p}{1-p}\right){}^{i}}.\label{eq:Comparison}
\end{align}

\begin{figure}[t]
\centering{}\includegraphics[width=7cm]{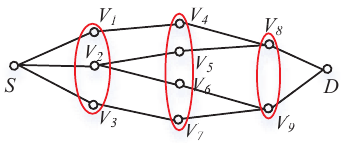}\caption{An illustration to show the malicious cut sets and minimal malicious
cut sets of a network. \label{Fig: illustration of cut sets} }
\end{figure}

Recall that both sub-matrix $\left[\begin{array}{ccc}
B_{1} & B_{s_{1}} & \mathbf{\boldsymbol{0}}\end{array}\right]$ and $\left[\begin{array}{ccc}
\mathbf{\boldsymbol{0}} & B_{s_{0}} & B_{0}\end{array}\right]$ correspond to a sub-network of the considered network and the common
nodes shared by the two sub-networks (if any) can not be malicious
vehicles. Therefore, when considering the potential malicious vehicle
combinations, we avoid these common nodes and only focus on the sub-matrix
$B_{1}$ and $B_{0}$. Specifically, we regard the network corresponding
to sub-matrix $B_{1}$ and $B_{0}$ as networks that each row represents
a complete path and each column represent a vehicle, denoted by $T_{1}$
and $T_{0}$ respectively. In the following, with a twist of the vertex-cut
\cite{Gross06} terminology from graph theory which defines a vertex
set whose removal would disconnect the graph, we define \textit{malicious
cut set}, \textit{size} of a malicious cut set, and \textit{minimal
malicious cut set} of a network in this paper, and demonstrate that
the parameter $a_{i}$, $1\leq i\leq n_{0}$ and $b_{i}$, $1\leq i\le n_{1}$
in \eqref{eq:Comparison}, which was defined in \eqref{definiton of a_i}
and \eqref{definiton of b_i}, are exactly the number of malicious
cut sets with size $i$ of the network $T_{0}$ and $T_{1}$ respectively.
\begin{defn}
\label{def:A-cut-set}A \textit{malicious cut set} of a network is
a combination of vehicles, where if all vehicles in the set are malicious
vehicles all paths of the network can be compromised. The \textit{size}
of a malicious cut set is the number of vehicles contained in the
set. A \textit{minimal malicious cut set }is a malicious cut set with
the smallest size. 
\end{defn}
It is worth noting that the network may have multiple malicious cut
sets and multiple minimal malicious cut sets. Consider the network
shown in Fig. \ref{Fig: illustration of cut sets} for an example.
Vehicle sets $\left\{ V_{1},V_{2},V_{3}\right\} $, $\left\{ V_{4},V_{5},V_{6},V_{7}\right\} $,
and $\left\{ V_{8},V_{9}\right\} $ (to name a few) are all malicious
cut sets of the network, and a minimal malicious cut set is the malicious
cut set $\left\{ V_{8},V_{9}\right\} $ with size $2$. Therefore,
to compromise all paths of this network, the minimum number of malicious
vehicles needed is $2$. 

Based on Definition \ref{def:A-cut-set}, if a vehicle set is a malicious
cut set, then each path of the network contains at least one vehicle
belongs to this set. Recall that $a_{i}$ (or $b_{i}$) represents
the number of combinations that randomly choosing $i$ columns from
sub-matrix $B_{0}$ ($B_{1})$ and then conducting element-wise union
to them, there results a column with each entry be $1$. That is,
$a_{i}$ (or $b_{i}$) represents the number of combinations that
by choosing $i$ vehicles from Network $T_{0}$ (or $T_{1}$) to form
a vehicle set, each path of network $T_{0}$ (or $T_{1}$) contains
at least one vehicle belongs to this set. Therefore, $a_{i}$, $1\leq i\leq n_{0}$
and $b_{i}$, $1\leq i\le n_{1}$ are exactly the number of malicious
cut sets with size $i$ of the network $T_{0}$ and $T_{1}$ respectively. 

According to the properties of malicious cut sets, it can be readily
obtained that $a_{i}=0$ if $a_{i+1}=0$, and $a_{i+1}>0$, if $a_{i}>0$.
Similarly, we have $b_{i}=0$ if $b_{i+1}=0$, and $b_{i+1}>0$, if
$b_{i}>0$.

Define
\begin{equation}
r_{0}=\min\left\{ i:\;a_{i}>0\right\} ,\;\;1\leq r_{0}\leq n_{0}\label{eq:definition of r0}
\end{equation}
and 
\begin{equation}
r_{1}=\min\left\{ i:\;b_{i}>0\right\} ,\;\;1\leq r_{1}\leq n_{1},\label{eq:definition of r1}
\end{equation}
the smallest integer that satisfies $a_{i}>0$ and $b_{i}>0$ respectively.
Therefore, $r_{0}$ is the size of the minimal malicious cut set of
network $T_{0}$, and $a_{r_{0}}$ is the number of minimal malicious
cut sets of network $T_{0}$. Similarly, $r_{1}$ is the size of the
minimal malicious cut set of network $T_{1}$, and $b_{r_{1}}$ is
the number of minimal malicious cut sets of network $T_{1}$. This
follows that
\begin{align}
\frac{\text{Pr}\left(M_{1}=m_{1},...M_{k}=m_{k}|m_{0}=1\right)}{\text{Pr}\left(M_{1}=m_{1},...M_{k}=m_{k}|m_{0}=0\right)}= & \frac{\sum_{i=r_{0}}^{n_{0}}a_{i}\cdot\left(\frac{p}{1-p}\right){}^{i}}{\sum_{i=r_{1}}^{n_{1}}b_{i}\cdot\left(\frac{p}{1-p}\right){}^{i}}\nonumber \\
\approx & \frac{a_{r_{0}}\left(\frac{p}{1-p}\right)^{r_{0}}}{b_{r_{1}}\left(\frac{p}{1-p}\right)^{r_{1}}},\label{eq:Approximate_Comparison}
\end{align}
where the first step is obtained from the fact that $a_{1}=a_{2}=...a_{r_{0}-1}=0$,
$a_{r_{0}}>0$, and $b_{1}=b_{2}=...b_{r_{1}-1}=0$, $b_{r_{1}}>0$,
and the second step is obtained by only keeping the first item of
both the numerator and denominator. Considering the fact that when
$p$ is small, the probability that there are $i+1$ malicious vehicle
in the network is much smaller than the probability that there are
$i$ malicious vehicles in the network, therefore, this approximation
is quite accurate. 

Note that when $p$ is small, we have $\frac{p}{1-p}\ll1$. Therefore,
when $r_{0}\neq r_{1}$, whether the value of $\frac{a_{r_{0}}\left(\frac{p}{1-p}\right)^{r_{0}}}{b_{r_{1}}\left(\frac{p}{1-p}\right)^{r_{1}}}$
shown as \eqref{eq:Approximate_Comparison} is larger than $1$ is
dominantly determined by the value of $r_{0}-r_{1}$. Specifically,
when $r_{0}<r_{1}$, we have $\left(\frac{p}{1-p}\right)^{r_{0}-r_{1}}\gg1$.
In this case, the coefficient $\frac{a_{r_{0}}}{b_{r_{1}}}$ plays
marginal role and therefore $\frac{a_{r_{0}}\left(\frac{p}{1-p}\right)^{r_{0}}}{b_{r_{1}}\left(\frac{p}{1-p}\right)^{r_{1}}}>1$;
when $r_{0}>r_{1}$, we have $\left(\frac{p}{1-p}\right)^{r_{0}-r_{1}}\ll1$,
and therefore $\frac{a_{r_{0}}\left(\frac{p}{1-p}\right)^{r_{0}}}{b_{r_{1}}\left(\frac{p}{1-p}\right)^{r_{1}}}<1$.
On the contrary, when $r_{0}=r_{1}$, whether the value of $\frac{a_{r_{0}}\left(\frac{p}{1-p}\right)^{r_{0}}}{b_{r_{1}}\left(\frac{p}{1-p}\right)^{r_{1}}}$
is larger than $1$ would heavily depend on the value of the coefficient
$\frac{a_{r_{0}}}{b_{r_{1}}}$. Consequently, we have
\begin{align}
\frac{\text{Pr}\left(M_{1}=m_{1},...M_{k}=m_{k}|m_{0}=1\right)}{\text{Pr}\left(M_{1}=m_{1},...M_{k}=m_{k}|m_{0}=0\right)} & \approx\frac{a_{r_{0}}\left(\frac{p}{1-p}\right)^{r_{0}}}{b_{r_{1}}\left(\frac{p}{1-p}\right)^{r_{1}}}\nonumber \\
 & \begin{cases}
>1, & r_{0}<r_{1}\\
<1, & r_{0}>r_{1}\\
=\frac{a_{r_{0}}}{b_{r_{1}}}, & r_{0}=r_{1}
\end{cases},\label{eq:Approximate_Compare_2}
\end{align}
which shows that to compare the values of $\text{Pr}\left(M_{1}=m_{1},...M_{k}=m_{k}|m_{0}=1\right)$
and $\text{Pr}\left(M_{1}=m_{1},...M_{k}=m_{k}|m_{0}=0\right)$, we
only need to compare the values of $r_{0}$ and $r_{1}$, namely,
the size of minimal malicious cut set of network $T_{0}$ and $T_{1}$
when $r_{0}\neq r_{1}$, or the value of $a_{r_{0}}$ and $b_{r_{1}}$,
namely, the number of minimal malicious cut sets of network $T_{0}$
and $T_{1}$ when they have the same size of minimal malicious cut
set.

From Menger's Theorem \cite{Gross06}, the size of the minimal vertex-cut
whose removal would disconnect two non-adjacent vertices, is equal
to the maximum number of vertex-independent paths between these two
non-adjacent vertices. Therefore, it can be concluded that the size
of minimal malicious cut set of a network is also equal to the maximum
number of node-disjoint paths in the network between the source vehicle
and the destination vehicle. Therefore, $r_{0}$ and $r_{1}$ are
also the numbers of maximum number of node-disjoint paths exist in
network $T_{0}$ and $T_{1}$ respectively. Note that calculating
the maximum number of vertex-disjoint paths from source to destination
is a special case of finding the maximum flow problem by setting every
vertex capacity $1$ \cite{Gross06}. Therefore, the values of $r_{0}$
and $r_{1}$ can be readily obtained by existing maximum flow algorithms,
e.g., introduced in \cite{Gross06,Karger96,Goldberg88}. When $r_{0}=r_{1}$,
$a_{r_{0}}$ and $b_{r_{1}}$ can be obtained by exhaustive search
algorithm according to their definitions given by \eqref{definiton of a_i}
and \eqref{definiton of b_i}. 

\begin{figure}[t]
\subfigure[Network 1.]{\label{eta_p_d}\includegraphics[width=4.3cm]{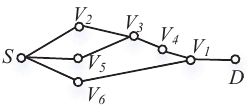}}\;\;\subfigure[Network 2.]{\label{eta_d_rho}\includegraphics[width=4.3cm]{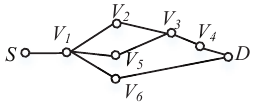}}

\caption{An illustration of two networks that have the same topology matrix.}

\label{Diff Networks}
\end{figure}

In summary, by combining \eqref{eq:MLE rule} and \eqref{eq:Approximate_Compare_2},
the decision rule of our proposed heuristic algorithm can be shown
as
\begin{equation}
d=\begin{cases}
1, & (r_{0}<r_{1})\;or\;(r_{0}=r_{1},\;a_{r_{0}}>b_{r_{1}})\\
0, & (r_{0}>r_{1})\;or\;(r_{0}=r_{1},\;a_{r_{0}}<b_{r_{1}})
\end{cases},\label{eq:Heuristic algorithm rule}
\end{equation}
and when $r_{0}=r_{1}$, and $a_{r_{0}}=b_{r_{1}}$, $d$ is randomly
chosen from $0$ and $1$ with equal probability. 
\begin{rem}
It is worth noting that in the above analysis, the network with a
topology matrix $B_{1}$ may not be unique. For instance, a topology
matrix $B=\left[\begin{array}{cccccc}
1 & 1 & 1 & 1 & 0 & 0\\
1 & 0 & 1 & 1 & 1 & 0\\
1 & 0 & 0 & 0 & 0 & 1
\end{array}\right]$ can correspond to both networks shown in Fig. \ref{Diff Networks}.
However, the malicious cut sets of the networks with different topology
remain the same as there is a one-to-one correspondence between each
malicious cut set and a combination of columns from the topology matrix
that an element-wise union of them resulting in a column with each
entry being 1. That is, as long as networks have the same topology
matrix $B$, they would have the same malicious cut sets. Therefore,
the network $T_{1}$ (or $T_{0}$) corresponding to the same sub-matrix
$B_{1}$ (or $B_{0}$) may not unique, however it does not affect
their malicious cut sets analysis. 
\end{rem}
\begin{rem}
The implication of the heuristic decision algorithm \eqref{eq:Heuristic algorithm rule}
can also be explained straightforwardly as follows. Given two networks
that deliver conflicting message contents, by removing the common
nodes shared by these two networks and regarding each path after the
removal of the common nodes as a new complete path, there results
in two new independent networks that deliver conflicting message contents.
Therefore, decision can be made by comparing the robustness of the
two new networks. Note that a smaller size of the minimal malicious
cut set of a network implies a less number of minimal malicious vehicles
are required to compromise that network, and consequently, a higher
probability to deliver incorrect messages. Therefore, the decision
will always be chosen as the message delivered by the network with
a lower probability to be compromised. 
\end{rem}
From \eqref{eq:Heuristic algorithm rule}, we can see that the decision
result is now entirely determined by the network topology, and is
independent of the proportion of malicious vehicles in the network.
That is, the proposed heuristic decision algorithm is purely topology-based
so that is easy to be implemented in practice. In summary, the heuristic
decision algorithm works as detailed in Algorithm \ref{alg:Heuristic-Decision-Algorithm}. 

\begin{algorithm}[h]
\caption{Heuristic Decision Algorithm\label{alg:Heuristic-Decision-Algorithm}}
\begin{lyxlist}{00.00.0000}
\item [{\textbf{INPUT}:}] $M_{1}...M_{k}$
\item [{\textbf{OUTPUT}:}] $d$
\item [{\textbf{begin}}]~
\end{lyxlist}
\begin{enumerate}
\item Construct topology matrix $B$ based on the paths information derived
from the received $k$ copies of message; 
\item Based on the constructed topology matrix $B$, calculate $r_{0}$
and $r_{1}$ based on maximum flow algorithm;
\item \textbf{If} $r_{0}<r_{1}$ \textbf{then} $d=1$\\
\textbf{}\\
\textbf{elseif} $r_{0}>r_{1}$ \textbf{then} $d=0$\\
\\
\textbf{else }calculate $a_{r_{0}}$ and $b_{r_{1}}$ based on their
definition given by \eqref{definiton of a_i} and \eqref{definiton of b_i};\textbf{
}\\
\textbf{}\\
\textbf{$\;\;\;\;$if $a_{r_{0}}>b_{r_{1}}$ then} $d=1$\textbf{}\\
\textbf{}\\
\textbf{$\;\;\;\;$elseif $a_{r_{0}}<b_{r_{1}}$ then} $d=0$\textbf{}\\
\textbf{}\\
\textbf{$\;\;\;\;$else $d$ }is randomly chosen from $0$ and $1$
with equal probability\textbf{}\\
\textbf{}\\
\textbf{$\;\;\;\;$end}\\
\textbf{}\\
\textbf{end}
\end{enumerate}
\begin{lyxlist}{00.00.0000}
\item [{\textbf{end}}]~
\end{lyxlist}
\end{algorithm}

\section{Simulation and Discussion\label{sec:Simulation-and-Discussion}}

In this section, we conduct simulations to establish the validity
of the decision algorithms proposed in Section \ref{sec:Optimum-Decision-Algorithm}
and Section \ref{sec:Heuristic-Decision-Algorithm}. We generate a
network that vehicles are Poissonly distributed in the road with density
$\rho$, and each relay vehicle has a probability $p$ to be a malicious
vehicle. Vehicles communicate with their neighbors adopting the unit
disk model \textcolor{black}{\cite{Mao09,Wang.Y16}} with transmission
range $r_{0}=250$m\textcolor{black}{{} }\cite{Zhang2014}. We focus
on a destination vehicle located at a distance $L$ from the source
vehicle. From the time instant the destination vehicle receives the
first message reporting road condition, it waits time period $T$
to receive more number of messages before it starts to make a decision.
The per-hop transmission delay is assumed to be $\beta=4$ms \cite{Zhang2014}.
For the road condition, we choose a rather conservative probability
of the occurrence of an abnormal situation. Specifically, we set that
an hazardous road/environmental condition happens randomly with probability
$0.001$ \cite{Raya08}, i.e., we set $P_{1}=\text{Pr}(m_{0}=1)=0.001$
and $P_{0}=1-P_{1}=0.999$. 

At each simulation, a topology matrix $B$ can be constructed based
on the underlying network topology. Therefore, given the malicious
vehicle distribution and the topology information, the content of
the $k$ messages $M_{1},M_{2},...M_{k}$ received by the destination
vehicle is determined. The destination vehicle then makes a decision
given the received messages and the derived underlying topology information
according to our proposed decision algorithms at each simulation.
The decision result can be either correct or incorrect. The simulation
is repeated 5000 times and the proportion of the correct decision,
i.e., the probability of correct decision, is plotted. 

In the following, we first compare our proposed two decision algorithms,
and then we study the effects of topology information, and some performance-impacting
parameters on the algorithms. The performance-impacting parameters
including the proportion of malicious vehicle in the network, the
choice of waiting time by the destination vehicle before it starts
to make the decision..

\subsection{Comparison of the two proposed algorithms}

In this part, we compare the message security performance achieved
by the two proposed decision algorithms to provide insight on the
optimum decision algorithm design for secure message dissemination. 

\begin{figure}[t]
\centering{}\includegraphics[width=8.5cm]{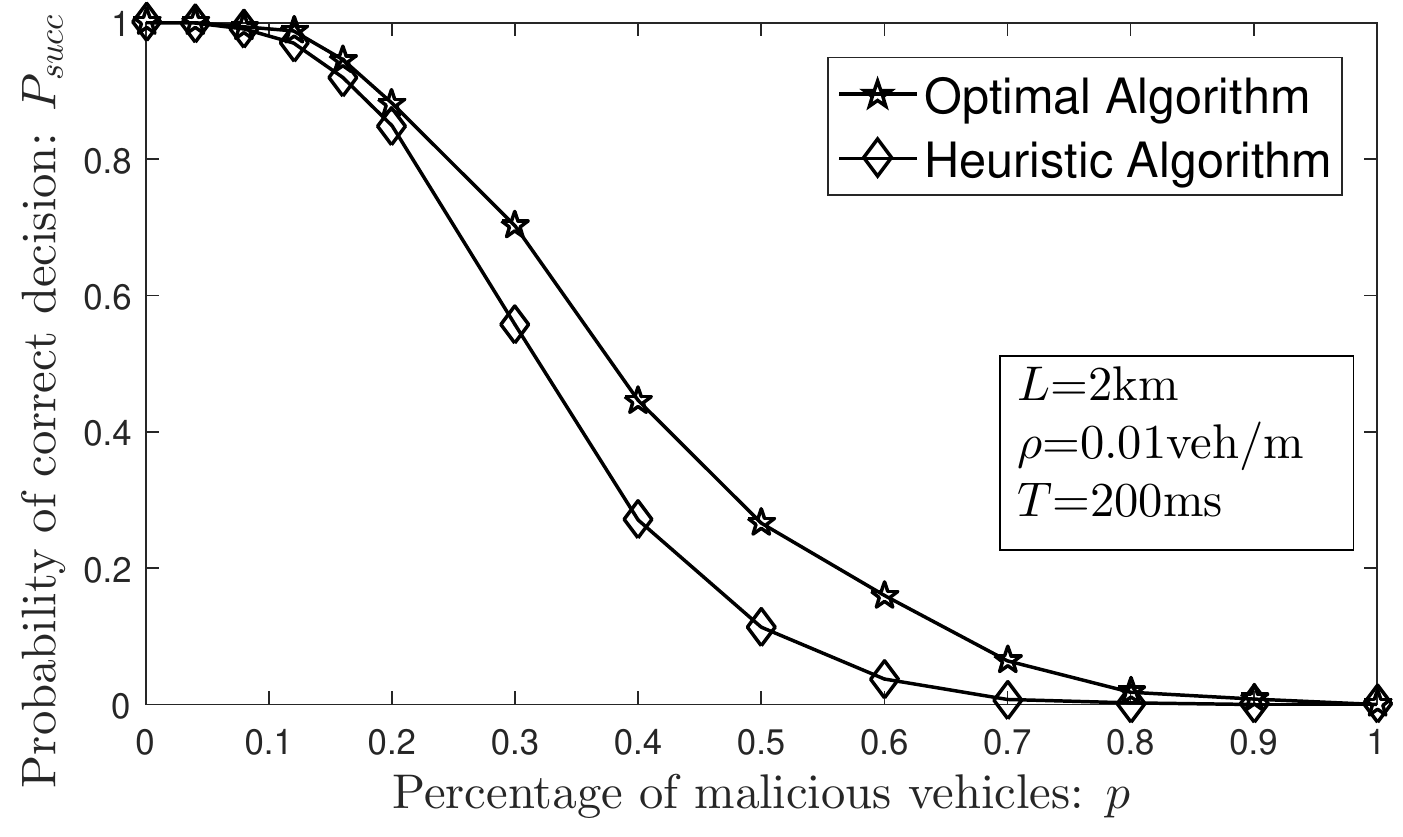}\caption{A comparison of the probabilities of correct decision achieved by
the optimum decision algorithm proposed in Section \ref{sec:Optimum-Decision-Algorithm},
and by the heuristic decision algorithm proposed in Section \ref{sec:Heuristic-Decision-Algorithm}.
\label{Fig: comparison of proposed algorithms} }
\end{figure}

Fig. \ref{Fig: comparison of proposed algorithms} compares the probability
of correct decision achieved by the proposed optimum decision algorithm
(labeled as Optimum Algorithm) and by the proposed pure topology-based
heuristic decision algorithm (labeled as Heuristic Algorithm) respectively.
It is shown that when the percentage of malicious vehicles in the
network is small, e.g., when $p<0.2$ in this case, the message security
performance achieved by the optimum decision algorithm is only slightly
better than the performance achieved by the heuristic decision algorithm.
This implies that the heuristic decision algorithm, that purely based
on network topology information and easily to be implemented in practice,
is sufficient to achieve a high message security performance for vehicular
networks. 

\subsection{Impact of topology information}

To evaluate the effectiveness of our proposed algorithms that takes
the underlying topology information into consideration, we compare
the security performance, in terms of the probability of correct decision
made by the destination vehicle, achieved by our proposed algorithms
described by Algorithm \ref{alg:Optimum-decision-algorithm} and \ref{alg:Heuristic-Decision-Algorithm}
respectively, with that achieved by existing weighted voting algorithms
like the weighted voting algorithm proposed in \cite{Zhu02} (labeled
with WV: MMSE) that considers partial correlation between messages,
the weighted voting algorithm proposed in \cite{Huang14} (labeled
with WV: $w\propto\alpha^{h-1})$ that does not consider the underlying
topology information causing the correlation between messages, and
the majority voting (a special case of weighted voting by assigning
identical weights to each vote) that totally ignores the underlying
topological correlation. Specifically, the weighted voting algorithm
proposed in \cite{Zhu02} set weight to each message as $w_{i}=\sum_{j=1}^{k}C_{ij}^{(-1)}\left(\sum_{r,j=1}^{k}C_{rj}^{(-1)}\right)^{-1}$,
where $C$ is the error covariance matrix whose $(i,j)$th entry is
defined by the error covariance between message $M_{i}$ and message
$M_{j}$, calculated by $C_{ij}=E\left[(M_{i}-m_{0})(M_{j}-m_{0})\right]$.
$C^{-1}$ is the inverse matrix of the error covariance matrix $C,$
and $C_{ij}^{(-1)}$ is the $(i,j)$th entry of the matrix $C^{-1}$.
The weighted voting algorithm proposed in \cite{Huang14} simply assigns
weight to each message as $w_{i}=\frac{\alpha^{h_{i}-1}}{\sum_{j}\alpha^{h_{j}-1}}$,
where $\alpha\in(0,1)$ is a weighting factor to reduce the oversampling
impact caused by messages generated from the same source and $h_{i}$
is the number of hops travelled by the $i$th message from the source
to the destination.

\begin{figure}[t]
\centering{}\includegraphics[width=8.5cm]{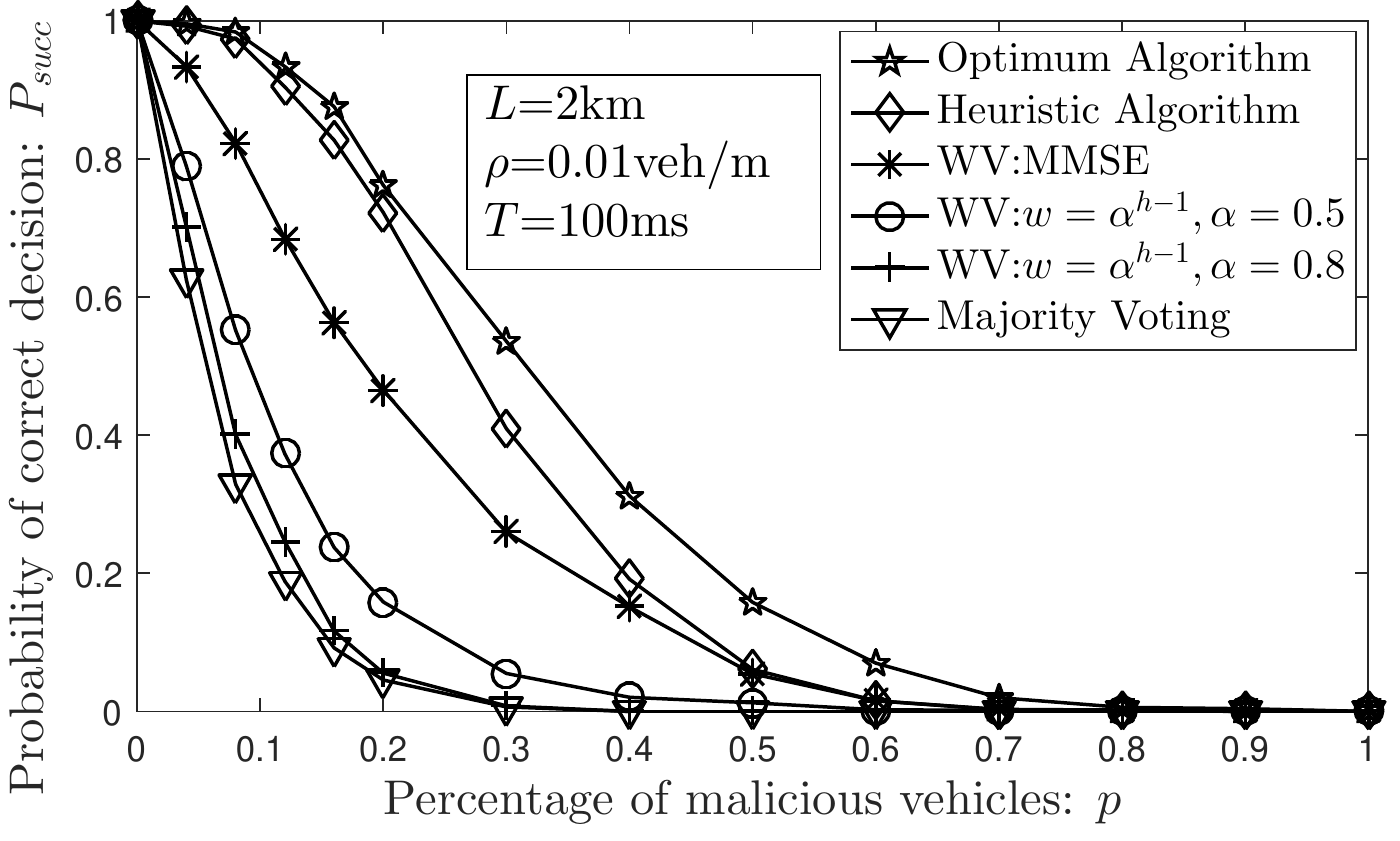} \caption{A comparison of the probability of correct decision achieved assuming
our proposed algorithms and that achieved assuming other existing
weighted voting algorithms. \label{Fig: Compare with existing algorithms} }
\end{figure}

It can be seen in Fig. \ref{Fig: Compare with existing algorithms}
that both our proposed algorithms outperform the weighted voting algorithms
proposed in \cite{Zhu02}, \cite{Huang14} and the majority voting
algorithm, which demonstrates that our algorithms taking into account
topology information and correlation between different copies of message
are able to effectively improve the robustness of vehicle networks
against attacks from malicious vehicles. 

\subsection{Impact of the percentage of malicious vehicles }

Fig. \ref{Fig: Compare with existing algorithms} reveals the relationship
between the probability of correct decision $P_{succ}$ and the percentage
of malicious vehicles in the network, $p$. It can be seen that the
probability of correct decision made by the destination vehicle decreases
to its minimum value $P_{succ}=0$ when the proportion of malicious
vehicles in the network is larger than a certain threshold. Beyond
that threshold, a further increase in $p$ has little impact on the
security performance. Specifically, as shown in Fig. \ref{Fig: Compare with existing algorithms},
when $p$ is small, the security performance achieved assuming the
optimum decision algorithm decreases with an increase of $p$; however,
when $p$ increases beyond a certain threshold, a further increase
in $p$ has no impact on the security performance. This can be explained
by the fact that the more malicious vehicles in the network, the more
tampered copies of message will be delivered, and therefore a lower
chance for the destination vehicle to make a correct decision regardless
of what algorithm it adopts. Furthermore, when the number of malicious
vehicles in the network reaches a certain threshold, most of the message
dissemination paths will be compromised. In this case, the destination
vehicle will totally misguided by the incorrect messages and the message
security performance approaches its minimum value $P_{succ}=0$. 

\subsection{Impact of the waiting time period }

As mentioned in Section \ref{subsec:Problem-formation}, the waiting
time period $T$ the destination vehicle waits before it starts to
make a decision is an important parameter that should balance the
trade-off between the response time requirement and the integrity
of the decision. Therefore, in this part, we study the impact of the
waiting time period $T$ on the security performance assuming the
two proposed algorithms, under different traffic densities.

\begin{figure}[t]
\centering{}\includegraphics[width=8.3cm]{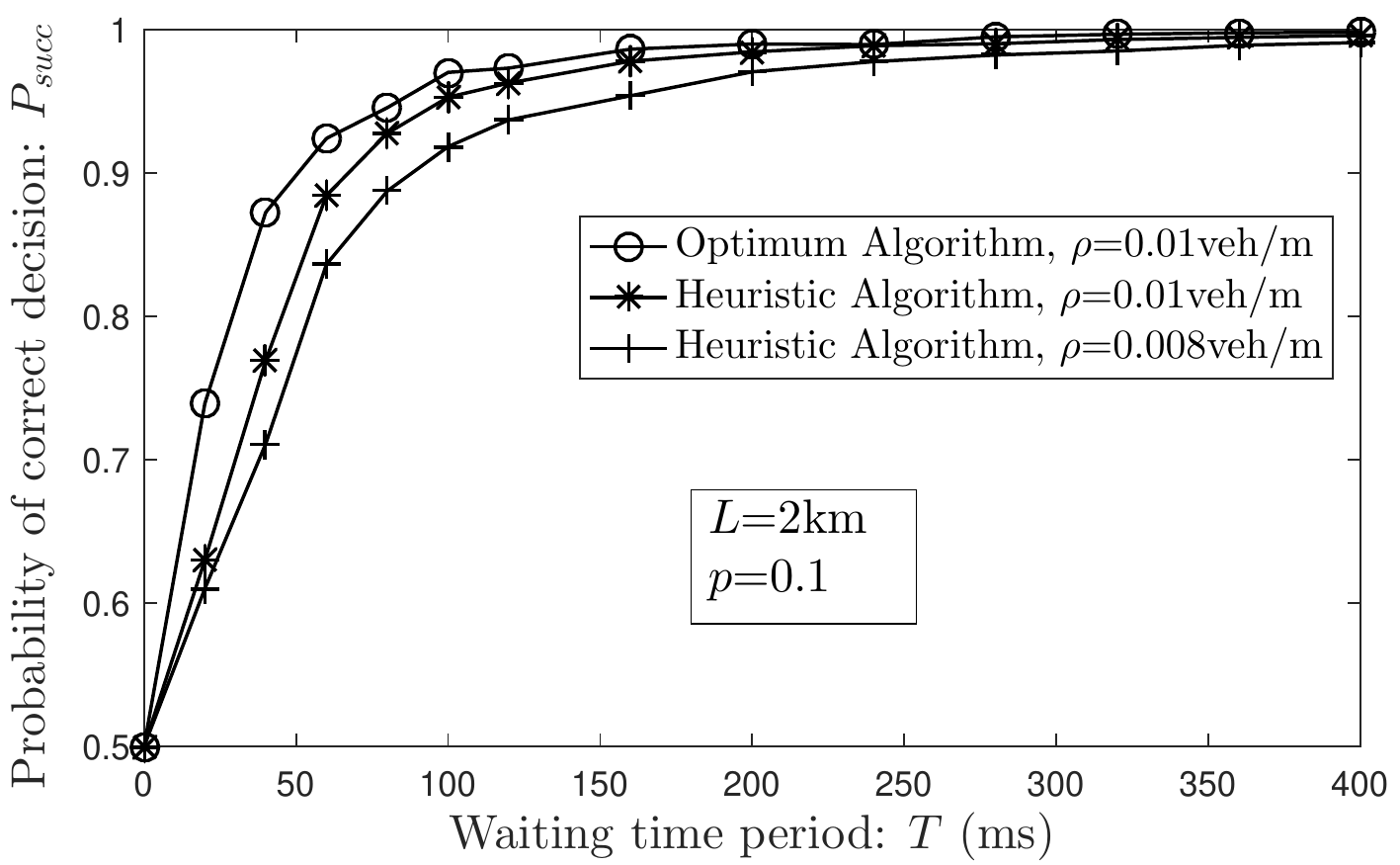} \caption{An illustration of the relationship between the probability of correct
decision and the waiting time period the destination vehicle waits
before it starts to make a final decision by adopting the proposed
two algorithms respectively. \label{fig:impact of number of messages}}
\end{figure}

Fig. \ref{fig:impact of number of messages} demonstrates the relationship
between the probability of correct decision, $P_{succ}$, and the
waiting time period $T$ the destination vehicle waits before it starts
to make a decision, assuming our two proposed algorithms respectively,
and gives insight into the choice of waiting time by the destination
vehicle.  Importantly, we can see that for both algorithms, a larger
number of waiting time is beneficial to the secure message dissemination
because a longer waiting time potentially implies a larger number
of received messages. This consequently, brings more information on
the underlying network topology, and therefore leads to a more robust
result of the data consistency check. However, when $T$ increases
beyond a certain threshold $T_{th}$, e.g., in the case of $\rho$=0.01veh/m,
$T_{th}=100ms$ when adopting the proposed optimum decision algorithm
and $T_{th}=150ms$ when adopting the proposed heuristic decision
algorithm when, a further increase in $T$ has marginal (less than
$5\%$) impact on the probability of correct decision. This is due
to the fact that when $T$ is larger than a threshold, the marginal
return brought by waiting a longer time to the security performance
is diminishing. Furthermore, it can be seen that to achieve the same
message security performance, when the vehicular density is lower,
the waiting time needs to be longer. Therefore, when determining the
waiting time period, it is important to take the vehicular density
into account, e.g., in areas where the vehicular density is large,
the waiting time can be reduced. Thus, Fig. \ref{fig:impact of number of messages}
exhibits a guide on the choice of waiting time period for destination
vehicles. 

\section{Conclusions \label{sec:Conclusion-and-Future}}

This paper proposed two decision algorithms that utilizes the underlying
network topology information to address the issue of message inconsistency
caused by malicious vehicles that would tamper the content of disseminated
messages. The optimum decision algorithm proposed is able to maximally
help a destination vehicle make a correct decision on the message
content, based on the network topology information and a prior knowledge
of the percentage of malicious vehicles in the network. The heuristic
decision algorithm proposed enables a vehicle to make a decision purely
based on network topology information, therefore is easier to implement
in practice. Simulations were conducted to verify the effectiveness
of the algorithms. We demonstrated that the heuristic decision algorithm
is able to achieve a security performance close to that achieved by
the optimum decision algorithm, especially when the percentage of
malicious vehicles in the network is small. By comparing the two proposed
algorithms with existing algorithms that do not consider the underlying
topological information or only partially consider message correlation,
we showed that our proposed algorithms greatly outperform existing
ones. Moreover, we discussed the impact of some key parameters on
the performance of the proposed algorithms, including the percentage
of malicious vehicles in the network, and the waiting time the destination
vehicle waits before making the final decision. Our results give insight
on the optimum decision algorithm design for vehicular networks to
improve message security. 

\begin{IEEEbiography}{Jieqiong Chen}
(S'16) received the Bachelor\textquoteright s degree in Engineering
from Zhejiang University, Zhejiang, China, in 2012, and she is currently
working toward the Ph.D. degree in engineering at the University of
Technology Sydney, NSW, Australia. Her research interests include
wireless communications and optimum vehicular network design for intelligent
transporatation systems. 
\end{IEEEbiography}

\begin{IEEEbiography}{Guoqiang Mao}
(S'98-M'02-SM'08-F'17) joined the University of Technology Sydney
in February 2014 as Professor of Wireless Networking and Director
of Center for Real-time Information Networks. Before that, he was
with the School of Electrical and Information Engineering, the University
of Sydney. He has published about 200 papers in international conferences
and journals, which have been cited more than 5000 times. He is an
editor of the IEEE Transactions on Wireless Communications (since
2014), IEEE Transactions on Vehicular Technology (since 2010) and
received \textquotedblleft Top Editor\textquotedblright{} award for
outstanding contributions to the IEEE Transactions on Vehicular Technology
in 2011, 2014 and 2015. He is a co-chair of IEEE Intelligent Transport
Systems Society Technical Committee on Communication Networks. He
has served as a chair, co-chair and TPC member in a large number of
international conferences. He is a fellow of IEEE and IET. His research
interest include intelligent transport systems, applied graph theory
and its applications in telecommunications, Internet of Things, wireless
sensor networks, wireless localization techniques and network performance
analysis. 
\end{IEEEbiography}

\begin{IEEEbiography}{Changle Li}
(M'09-{}-SM'16) received the Ph.D. degrees in communication and information
system from Xidian University, China, in 2005. Since then, he conducted
his postdoctoral research in Canada and the National Institute of
information and Communications Technology (NICT), Japan, respectively.
He has been a visiting scholar at the University of Technology Sydney
(UTS) and is currently a Professor with the State Key Laboratory of
Integrated Services Networks, Xidian University. He is an IEEE Senior
Member and his research interests include intelligent transportation
systems, vehicular networks, mobile ad hoc networks, and wireless
sensor networks. 
\end{IEEEbiography}

\begin{IEEEbiography}{De-gan Zhang}
(M\textquoteright 01) Born in 1970, Ph.D. Graduated from Northeastern
University, China. Now he is professor of Tianjin Key Lab of Intelligent
Computing and Novel software Technology, Key Lab of Computer Vision
and System, Ministry of Education, Tianjin University of Technology,
Tianjin, 300384, China. His research interest includes IOT, WSN, IOV,
etc. His E-mail: gandegande @126.com. He is the corresponding author
of this paper. 
\end{IEEEbiography}

\end{document}